\documentstyle[preprint,prl,aps]{revtex}
\begin{document}
\draft
\title{Asymptotic Bethe-Ansatz Results
for a Hubbard Chain with 1/sinh-Hopping}

\author{P.-A.\ Bares$^{\rm (a)}$ and F.\ Gebhard$^{\rm (a,b)}$}
\address{$^{(a)}$
Institut Laue-Langevin, B.P.~156x, F-38042 Grenoble Cedex, France}
\address{$^{(b)}$ Dept.~of
Physics and Materials Sciences Center, Philipps University Marburg,
D-35032~Marburg, Germany}

\maketitle%
\begin{abstract}%
We investigate spin-1/2 electrons with local Hubbard interaction
and variable range hopping amplitudes
which decay like~$\sinh(\kappa)/\sinh(\kappa r)$.
Assuming integrability the Asymptotic Bethe Ansatz
approach allows us to derive the generalized Lieb-Wu integral equations from
the two-particle phase shift.
Due to the nesting property there is a metal-to-insulator transition
at $U_c(\kappa>0)=0^+$. The charge gap in the singular
limit $\kappa=0$ opens when the interaction strength equals the bandwidth,
$U_c(\kappa=0)=W >0$.
\end{abstract}


Exact solutions play a crucial role in our understanding of strongly
correlated systems.
The algebraic and the
coordinate Bethe Ansatz constitute the main line of approach to integrable
systems~\cite{Shastrybook}.
Nevertheless, there are exactly solvable Hamiltonians
whose eigenstates cannot be cast into the standard Bethe Ansatz form.
A particular example is the popular Calogero-Sutherland-Moser
Hamiltonian~\cite{Calogero,SutherlandABA}
that has attracted recently a great deal
of attention in connection with the
universal properties of disordered systems and
random matrix theory~\cite{Simonsetal,Zirnbauer}.

Recently, Ruckenstein and one of us (F.G.)
introduced a $1/r$-Hubbard model that
incorporates long-range hopping and on-site repulsion, and that includes
the lattice version of the Calogero-Sutherland-Moser model in the
strong-coupling limit at half band-filling~\cite{prlgr}.
The integrability
as well as the structure of the eigenfunctions of this model are still
poorly understood.
The analysis of the two-body problem
in the $1/r$-Hubbard model indicates that the eigenstates are nei\-ther of the
Bethe Ansatz form as in the Hubbard model,
nor of the Jastrow-type as in the Calogero-Sutherland
model~\cite{GebhardHabil}.
Since the wave functions do not become plane waves even at large
distances, Sutherland's Asymptotic Bethe
Ansatz~\cite{Shastrybook,SutherlandABA} cannot
be directly applied to the $1/r$-Hubbard model.

In this article we introduce a model with variable range hopping
that interpolates between the standard and the $1/r$-Hubbard model.
For half-filled bands and in the strong coupling limit it reduces to
the antiferromagnetic $1/\sinh^2(\kappa r)$-Heisenberg
or Inozemtsev model~\cite{Inozemtsev}. This model is a
special case of an exchange interaction model~\cite{PolySS} most
recently explicitly solved in Ref.~\cite{SRS}.
The $1/\sinh(\kappa r)$-Hubbard model is a straightforward but
non-trivial generalization of the Inozemtsev model to
an itinerant electron system~\cite{Alex}.

We consider a one-dimensional
Hubbard model~\cite{Hubbard} for~$N$ spin-1/2 electrons
with hopping amplitudes~$t(l-m)$ and on-site repulsion~$U$,
\begin{equation}
\hat{H} = \sum_{l \neq m,\sigma} t(l-m)\hat{c}_{l,\sigma}^{+}
\hat{c}_{m,\sigma}^{\mbox{}}
+ U \sum_{l} \hat{n}_{l,\uparrow}\hat{n}_{l,\downarrow}  \; ,
\label{hamilt}
\end{equation}
where the lattice sums on the ring
run from $-L/2$ to $L/2-1$.
The Hubbard model~(\ref{hamilt}) is known
to be exactly solvable in two cases:
(i)~$t (l-m)=-t \delta_{l-m,\pm1}$ for the
standard Hubbard model~\cite{LiebWu}
and (ii)~$t(l-m)=-it (-1)^{l-m}[d(l-m)]^{-1}$
for the $1/r$-Hubbard model~\cite{prlgr}
where $d(l-m)= (L/\pi) \sin[\pi (l-m)/L]$ denotes the chord distance
between the sites $l$ and $m$. Henceforth we will use~$t$ as our energy unit.
Here we introduce a model on the infinite chain ($L \to\infty$)
with hopping amplitudes
$t (l-m)= -i \sinh (\kappa) (-1)^{l-m} /\sinh [\kappa(l-m)]$
where $\kappa^{-1}$ controls the effective range of the hopping.

The dispersion relation $\epsilon (k)$ of our model~(\ref{hamilt})
is given by $\epsilon (k)=(-2i)\sum_{n=1}^{\infty} t(n) \sin (k n)$.
It is odd under parity,
and can be expressed in terms of a logarithmic derivative of theta-functions.
For $\kappa=0$ one has $\epsilon (k) =k$ with a discontinuity at
the Brillouin zone boundary, while for $\kappa\to\infty$
one finds $\epsilon (k)= 2 \sin k$.
$\epsilon(k)$ is continuous for all~$\kappa>0$
with zeros at $k=\pi$, $k=0$. The maximum (minimum)
at  $k= \pi/2$ ($k= -\pi/2$) for $\kappa=\infty$ is gradually shifted
as function of~$\kappa$ to higher (lower) momenta
until it reaches~$k=\pi$ ($k=-\pi$) for~$\kappa=0$.

The corresponding low-energy ($g$-ology~\cite{Solyom})
Hamiltonian involves left- and right-movers
with {\em different} velocities.
The limit $\kappa\to 0$ is singular as the dispersion relation involves
suddenly only right-movers,
while the left-movers' velocity diverges like~$1/\kappa$, i.e.,
a discontinuity occurs in~$\epsilon (k)$ at the Brillouin zone boundary.
Hence the physics of the metal-to-insulator transition
for $\kappa >0$ and for $\kappa=0$ ($1/r$-Hubbard model)
is completely different.
At half-filling  the distance
$2k_F=\pi n$ between the two Fermi points becomes half a reciprocal
lattice vector (``perfect nesting'').
The bands for right- and left-movers effectively
cross each other, and their degeneracy is lifted for all $U >0$.
The critical value for the transition is thus $U_c(\kappa >0)
=0^+$. In the chiral $1/r$-Hubbard model where left-movers are absent
one obtains a finite critical value,
$U_c(\kappa=0)=2\pi $, see Ref.~\cite{prlgr}.

We discuss the symmetric orbital part $\psi(x_1,x_2)$ of the two-particle
wave function in the limit of a large system
with open boundary conditions.
In first quantization
the Schr\"{o}dinger equation for the wave function
$\psi(x_1,x_2)$ reads for $x_1 \neq x_2$
\label{Sgl}
\begin{equation}
E \psi(x_1,x_2)  =  F(x_1,x_2) + F(x_2,x_1)
+ t(x_1-x_2) \left( \psi(x_2,x_2) -\psi(x_1,x_1) \right) \; ,
\end{equation}
while for $x_1=x_2$ we find $(E-U) \psi(x_1,x_1) = 2 F(x_1,x_1)$.
Here we defined
$F(x_1,x_2) =  \sum_{x\neq x_1, x_2} t(x_1-x) \psi(x,x_2)$.

We seek symmetric, spin singlet
scattering solutions 
$\psi(x_1 \ll x_2) = e^{i(k_1 x_1+k_2 x_2)} - e^{i\theta(k_1,k_2)}
 e^{i(k_2 x_1+k_1 x_2)}$,
where $\theta(k_1,k_2)$ is the phase shift,
and $k_1$, $k_2$ are the quasi-momenta.
Furthermore, we should choose
$\psi(x_1,x_2)$ such that we can employ the trigonometric identity
$\sinh(z_1) \sinh(z_2) = \sinh(z_1-z_2)/\left[\coth(z_2) - \coth(z_1)\right]$.
This naturally leads to
\begin{eqnarray}
\psi(x_1,x_2) &=& \frac{ 1-\delta_{x_1,x_2}}{2 \sinh[\kappa(x_2-x_1)]}
\biggl[ e^{\kappa(x_2-x_1)}
\left( A e^{i(k_1x_1+k_2x_2)} - B e^{i(k_2x_1+k_1x_2)} \right)
\nonumber\\[6pt]
&& \phantom{\biggl[} - e^{-\kappa(x_2-x_1)}
\left( A e^{i(k_1x_2+k_2x_1)} - B e^{i(k_1x_1+k_2x_2)} \right)
\biggr] + \delta_{x_1,x_2} \lambda e^{i (k_1+k_2)x_1}
\label{wave_function}
\end{eqnarray}
as choice for the wave function ($B/A=e^{i\theta(k_1,k_2)}$).
It has precisely the form of Ino\-zem\-tsev's two-magnon
state~\cite{Inozemtsev}.

The calculation of~$F(x_1,x_2)$ has to be done with care because of
the infinite lattice sums.
For $x_1\neq x_2$ the Schr\"{o}dinger equation gives
$E =\epsilon(k_1)+\epsilon(k_2)$, and
\begin{equation}
\lambda = (A-B) + \frac{(A+B)(\epsilon(k_1)-\epsilon(k_2))}{2i \sinh
(\kappa)} \quad ,
\end{equation}
while the equation for $x_1=x_2$ becomes
$(A-B) E -i(A+B) (\eta(k_1)-\eta(k_2)) = \lambda (E-U) $
with the abbreviation
$\eta (k) = 2\sinh(\kappa) \sum_{n=1}^{\infty}
(-1)^{n} \cos(k n) \cosh(\kappa n)/\sinh^2 (\kappa n) = \eta(-k)$.
The phase shift is then found as
\begin{equation}
\theta(k_1,k_2) = -2 \tan^{-1} \left[ \frac{H(k_2)-H(k_1)}{U/2}\right]
\label{phaseshift}
\end{equation}
with $H(k)=\left[ -\eta (k) +\epsilon(k)
(\epsilon (k)-U)/(2 \sinh(\kappa))
\right]/2$.

Sutherland~\cite{Shastrybook,SutherlandABA} observed that the two-particle
phase shift is sufficient to obtain the (scattering) spectrum
of a model, even if the wave functions cannot explicitly be constructed
as long as the model is integrable.
It is far from clear that the model~(\ref{hamilt}) is integrable.
Nonetheless we investigate the consequences of this conjecture.

We employ periodic boundary conditions to quantize
the pseudo momenta~$k_i$. The error introduced here is exponentially
small, of order $\exp (-\kappa L)$.
It is clear how to set up the Asymptotic Bethe Ansatz equations
for~$N-M$ up-spins and $M$ down-spins in a straightforward generalization
of the Lieb-Wu equations~\cite{LiebWu} because~$H(k)$ plays the role
of~$\sin k$:
\begin{mathletters}
\label{LiebWu}
\begin{eqnarray}
 L k_j  &=&  2\pi I_j + \sum_{\beta=1}^{M} \Theta\left( 2 H(k_j) -2
\Lambda_{\beta} \right)  \\
 - \sum_{j=1}^{N} \Theta \left( 2 \Lambda_{\alpha} - 2 H(k_j) \right)
 &=& 2\pi J_{\alpha} - \sum_{\beta=1}^{M} \Theta\left( \Lambda_{\alpha}-
 \Lambda_{\beta} \right)
\end{eqnarray}
\end{mathletters}%
for $ j=1,\ldots\ ,N$; $ \alpha=1,\ldots\ ,M$,
where $\Theta (x)=-2 \tan^{-1}\left(2x/U\right)$.
$I_j$ are integers (half-odd integers) for $M$ even (odd),
$J_{\alpha}$ are integers (half-odd integers) for $N-M$ even (odd).

We are interested in the ground state energy per site, $E/L$, of the model
in the thermodynamic limit. In this case the equations~(\ref{LiebWu})
can be transformed into integral equations for the densities
$ \rho(k) = 1/\left[ L(k_{j+1}-k_j)\right]$,
$\sigma(\Lambda) = 1/\left[ (\Lambda_{\alpha+1}-\Lambda_{\alpha})\right]$
in the usual way:
\begin{mathletters}
\begin{eqnarray}
2\pi \rho(k) &=& 1 \!+\! H'(k) \int_{B_1}^{B_2} d\lambda \sigma(\lambda)
 K\left[ H(k)-\lambda; U/4 \right] \\[3pt]
2\pi \sigma(\lambda) &=& \int_{Q_1}^{Q_2} dk \rho(k)
 K\left[ H(k)-\lambda; U/4 \right]
- \int_{B_1}^{B_2} d \lambda' \sigma(\lambda')
 K\left[ \lambda-\lambda'; U/2 \right]
\end{eqnarray}%
\end{mathletters}%
with $K\left[x;y\right] = 2y/(x^2+y^2)$, and $H'(k)=
\partial H(k)/\partial k$.
The ground state energy density is calculated as
$E/L  =  \int_{Q_1}^{Q_2} dk \rho(k) \epsilon(k)$.

The integration limits ~$Q_1$, $Q_2$, $B_1$, and $B_2$
are no longer symmetric around zero. They are determined
by the particle numbers,
$\int_{Q_1}^{Q_2} dk \rho(k)  = N/L$,
$\int_{B_1}^{B_2}  d\lambda \sigma (\lambda) = M/L$,
and the condition that a charge (spin)
excitation at $Q_1$ ($B_1$) has the same energy as the corresponding
excitation at $Q_2$ ($B_2$). The latter conditions are most conveniently
described in the pseudo particle picture~\cite{FrahmKorepin}.
The pseudo particle dispersions for charge and spin
for zero external magnetic field follow from
\begin{mathletters}
\begin{eqnarray}
\epsilon_c (k) &=& \epsilon(k) + \int_{B_1}^{B_2}
d\lambda \epsilon_s(\lambda)
K\left[ H(k)-\lambda; U/4 \right] \\[3pt]
2\pi \epsilon_s(\lambda)  &=& \int_{Q_1}^{Q_2} dk H'(k) \epsilon_c(k)
K\left[ H(k)-\lambda; U/4 \right]
 - \int_{B_1}^{B_2} d \lambda' \epsilon_s(\lambda')
K\left[ \lambda-\lambda'; U/2 \right] \; .
\end{eqnarray}%
\end{mathletters}%
The integration bounds have to fulfill
$\epsilon_c(Q_1) = \epsilon_c(Q_2)$,
$\epsilon_s(B_1) = \epsilon_s(B_2)$.
The ground state energy and chemical potential
can then be expressed in terms of the pseudo particle energies as
$E/L = \int_{Q_1}^{Q_2}  dk \epsilon_c(k)/(2\pi)$, and
$\mu = \mathop{{\rm Max}}_{Q_1 \leq k \leq Q_2} \epsilon_c(k)$.

For zero external magnetic field, $M=L/2$, one finds $B_2=\infty=-B_1$.
For half-filling, $N=L$, one further obtains $Q_2=\pi=-Q_1$.
Therefore we can solve the integral equations
analytically using Fourier transformation. With the help
of the four functions
\begin{eqnarray}
\widetilde{J}_0^{c \choose s}(\omega)&=& {{\rm Re} \choose {\rm Im}}
\int_{-\pi}^{\pi}
\frac{dk}{2\pi} e^{i\omega H(k)} \nonumber \\[6pt]
\widetilde{J}_1^{c \choose s}(\omega)&=& {{\rm Re} \choose {\rm Im}}
\int_{-\pi}^{\pi}
\frac{dk}{2\pi} e^{i\omega H(k)} \epsilon(k) H'(k) \quad . \nonumber
\end{eqnarray}
one may express the densities and the ground state energy as
\begin{mathletters}
\begin{eqnarray}
\sigma (\lambda) &=&
\int_{0}^{\infty} \frac{d\omega}{2\pi \cosh(\omega U/4)} \Bigl[
 \cos (\omega\lambda) \widetilde{J}_0^{c}(\omega)
+ \sin (\omega\lambda) \widetilde{J}_0^{s}(\omega) \Bigr] \\[6pt]
\rho(k) &=& \frac{1}{2\pi} + H'(k)
\int_{0}^{\infty} \frac{d\omega}{\pi(1+ e^{\omega U/2})} \Bigl[
 \cos (\omega H(k)) \widetilde{J}_0^{c}(\omega)
+ \sin (\omega H(k)) \widetilde{J}_0^{s}(\omega) \Bigr] \\[6pt]
E/L &=&  \int_0^{\infty} \frac{2\, d\omega}{1+e^{\omega U/2}}
\left[ \widetilde{J}_0^c(\omega)\widetilde{J}_1^c(\omega)
+ \widetilde{J}_0^s(\omega)\widetilde{J}_1^s(\omega)\right] \; .
\label{ezero}
\end{eqnarray}
\end{mathletters}%
For $\kappa\to\infty$ we have $\widetilde{J}^s_0(\omega)=0$,
$\widetilde{J}^c_0(\omega)=J_0(\omega)$,
$\widetilde{J}^c_1(\omega)=-2 J_1(\omega)/\omega$, where $J_0$, $J_1$
are Bessel functions, and eq.~(\ref{ezero})
becomes the Lieb-Wu result for the ground state energy~\cite{LiebWu}.
For~$\kappa\to 0$ one may rescale
$\omega= z/\kappa$ which allows to perform the calculation
analytically. The final result is $E/L(U\leq W)=(-W+U)/4-U^2/(12W)$,
$E/L(U\geq W)=-W^2/(12U)$,
with the band-width $W=2\pi$,
in complete agreement with Ref.~\cite{prlgr}.
$E/L$ is monotonically increasing as function of~$U/t$
and shows no singular behavior as function of~$\kappa$.

The pseudo particle dispersions become
\begin{eqnarray}
\epsilon_s (\lambda) &=&
\int_{0}^{\infty} \frac{d\omega}{\cosh(\omega U/4)}
 \Bigl[ \cos (\omega \lambda) \widetilde{J}_1^c(\omega)
+ \sin (\omega \lambda) \widetilde{J}_1^s(\omega) \Bigr] \nonumber
\\[6pt]
\epsilon_c(k) &=& \epsilon(k) +
2 \int_{0}^{\infty} \frac{d\omega}{1+ e^{\omega U/2}} \Bigl[
 \cos \left[ \omega H(k) \right] \widetilde{J}_1^c(\omega)
+ \sin \left[ \omega H(k) \right] \widetilde{J}_1^s(\omega) \Bigr]
\; .
\label{epsc}
\end{eqnarray}
The gap $\Delta\mu = \mu(n\to 1^+)- \mu(n\to1^-)
= U-2 \mu(n\to1^-)$ is given by
$\Delta\mu= U-2 {\rm Max}(\epsilon_c(k))$.
For large~$U/W$ one finds $\Delta \mu(U\gg W) = U-W +{\cal O} (W^2/U)$.
The analytical structure of~$\Delta\mu$ is very
similar for all~$\kappa >0$ such that the physics cannot be different from
$\kappa=\infty$ where $U_c=0^+$. Consequently, $U_c(\kappa >0)=0^+$,
and the gap is exponentially small for $U \ll W$.
The situation changes for~$\kappa=0$. Explicitly, $\epsilon_c(k,\kappa=0)=
\left[ 2 k +U-\sqrt{U^2+W^2-2WUk/\pi}\, \right]/4$
such that $\Delta\mu=0$ for $U\leq W$, and
$\Delta\mu=U-W$ for $U\geq W$, i.e., $U_c(\kappa=0)=W$, in agreement
with Ref.~\cite{prlgr} and the discussion above.

The ground state energy for strong coupling and half-filling
can be cast into the form
($2\sinh(\kappa)A=W/2$; $\psi(x)$ is the digamma function)
\begin{mathletters}
\label{largeulimit}
\begin{eqnarray}
E/L &=& -\frac{1}{U} \int_{-A}^{A} \frac{dx}{2\pi} e(x)
\int_{-A}^{A} dx' \rho(x)
{\rm Re } \left[ \psi\left(\frac{1}{2}+i \frac{x-x'}{2} \right)
- \psi\left(1+i \frac{x-x'}{2} \right) \right] \\[6pt]
e(x)& =& 2\sinh(\kappa) \left[ \eta\left( \epsilon_+^{-1}
(2x \sinh(\kappa)) \right)
- \eta\left( \epsilon_-^{-1} (2x \sinh(\kappa)) \right)
\right] \\[6pt]
\rho(x)& =& \frac{\sinh(\kappa)}{\pi} \left[
\frac{1}{\epsilon'\left( \epsilon_+^{-1} (2x \sinh(\kappa))\right)}
- \frac{1}{\epsilon'\left( \epsilon_-^{-1} (2x \sinh(\kappa))\right)}
\right]  \; .
\end{eqnarray}
\end{mathletters}%
Here, $k_{\pm}=\epsilon_{\pm}^{-1} (y)$ are the two solutions of
$\epsilon(k)=y$ with $\epsilon'(k_+) >0$, \hbox{$\epsilon'(k_-)<0$}.
Numerical results for $\kappa=\infty; 1; 0.1; 0$ are
$\lim_{U\to\infty} (U E/L)=
-4 \ln 2 \approx -2.773; -2.826; -3.197; -\pi^2/3 \approx -3.290$,
respectively.
The expressions~(\ref{largeulimit}) agree with the result obtained
by Sutherland et al.~\cite{SRS} which supports our
assumptions about integrability of the model~(\ref{hamilt}).
They also provide an explicit solution
for the integral equations in Ref.~\cite{SRS} for the Inozemtsev model.
We have calculated the critical exponents that control the asymptotic
behavior of correlation functions by finite size scaling in conformal
theory~\cite{FrahmKorepin}. The final formulae generalize those of
Ref.~\cite{FrahmKorepin} to the case of different velocities
for right- and left-moving electrons~\cite{GebhardBares}.

In sum, we are confident that the model~(\ref{hamilt}) is integrable
and can thus be solved using the Asymptotic Bethe Ansatz.
A complete construction of scattering states at {\em finite} densities
is still lacking even for the $1/\sinh^2(\kappa r)$-Heisenberg
model~\cite{Inozemtsev}. Models with variable-range exchange thus remain
an interesting problem in mathematical physics.


We thank P.~Nozi\`{e}res and A.~Gogolin for
stimulating discussions. F.~G.\ thanks all his colleagues at
the ILL~Grenoble for their kind hospitality during a stay in~1993/94.

\end{document}